\documentclass[a4paper,UKenglish,cleveref, autoref, thm-restate]{lipics-v2021}
\usepackage{appendix} 
\title{RCM++:Reverse Cuthill-McKee ordering with Bi-Criteria Node Finder} 

\author{JiaJun HOU\footnote{Corresponding author}}{Guangdong Provincial Key Laboratory of Interdisciplinary Research and Application for Data Science, BNU-HKBU United International College, China}{r130026044@mail.uic.edu.cn}{}{}
\author{HongJie Liu}{Guangdong Provincial Key Laboratory of Interdisciplinary Research and Application for Data Science, BNU-HKBU United International College, China}{r130033019@mail.uic.edu.cn}{}{}
\author{ShengXin Zhu}{Research Center for Mathematics, Beijing Normal University, China\\
Guangdong Provincial Key Laboratory of Interdisciplinary Research and Application for Data Science, BNU-HKBU United International College, China}{Shengxin.Zhu@bnu.edu.cn}{}{}

\authorrunning{J. HOU, H. Liu, and S. Zhu}

\Copyright{JiaJun HOU, HongJie Liu, S. Zhu} 



\nolinenumbers 

\EventEditors{John Q. Open and Joan R. Access}
\EventNoEds{2}
\EventLongTitle{42nd Conference on Very Important Topics (CVIT 2024)}
\EventShortTitle{CVIT 2024}
\EventAcronym{CVIT}
\EventYear{2024}
\EventDate{ 4/10, 2024}
\EventLocation{Little Whinging, United Kingdom}
\EventLogo{}
\SeriesVolume{42}
\ArticleNo{23}

\begin{document}
    
\maketitle

\begin{abstract}

The Reverse Cuthill-McKee (RCM) algorithm \cite{chan1980linear} is a graph-based method for reordering sparse matrices, renowned for its effectiveness in minimizing matrix bandwidth and profile. This reordering enhances the efficiency of matrix operations, making RCM pivotal among reordering algorithms. 
In the context of executing the RCM algorithm, it is often necessary to select a starting node from the graph representation of the matrix. This selection allows the execution of BFS (Breadth-First Search) to construct the level structure. The choice of this starting node significantly impacts the algorithm's performance, necessitating a heuristic approach to identify an optimal starting node, commonly referred to as the RCM starting node problem. Techniques such as the minimum degree method and George-Liu (GL) \cite{george1981solution} algorithm are popular solutions.

This paper introduces a novel algorithm addressing the RCM starting node problem by considering both the eccentricity and the width of the node during the run. Integrating this algorithm with the RCM algorithm, we introduce RCM++. Experimental results demonstrate that RCM++ outperforms existing RCM methods in major software libraries, achieving higher quality results with comparable computation time. This advancement fosters the further application and development of the RCM algorithm.The code related to this research has been made available at \href{https://github.com/SStan1/RCM_PP.git}{https://github.com/SStan1/RCM\_PP.git}.

\end{abstract}

\begin{figure*}[!htb]
\centering
  \subfloat[Original Matrix]{
    \includegraphics[width=7cm]{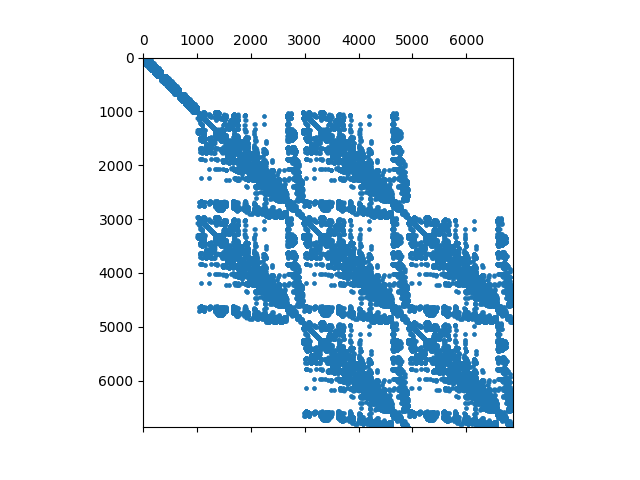}}
  \subfloat[Matrix with RCM reordering]{
    \includegraphics[width=7cm]{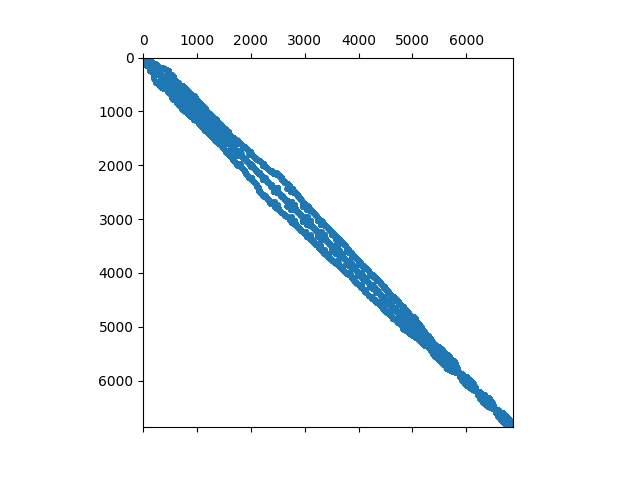}}
  \caption{The figure used to demonstrate the effect of the Reverse Cuthill-McKee algorithm.}
\end{figure*}

\baselineskip=18pt plus.2pt minus.2pt
\parskip=0pt plus.2pt minus0.2pt

\section{Introduction}

Matrix reordering is a well-established technique for sparse matrix preprocessing. By reordering the rows and columns of a matrix, this technique can give the matrix a more regular, computer-friendly shape.

In this paper, we focus on the Reverse Cuthill-McKee (RCM) algorithm, a prominent matrix reordering technique. RCM reorders a sparse matrix to concentrate non-zero elements near the diagonal which creates a banded structure. This reordering improves cache efficiency by increasing cache hits and reducing slow memory fetches, significantly accelerating matrix operations and overall computational efficiency.
Its simplicity and adaptability have led to widespread application in fields such as graph theory, parallel computing, and finite element analysis. 

In their study, Trotter et al. \cite{10.1145/3581784.3607046} highlight the RCM algorithm's  exceptional capability in reducing matrix profile and bandwidth. Such attributes make the RCM algorithm particularly advantageous in modern computer architectures, where data localization is crucial for computational performance.For instance, the RACE coloring engine \cite{alappat2020recursive} employs RCM to preprocess matrices before conducting parallel computing, thus streamlining subsequent computational processes. These observations underscore the RCM algorithm's pivotal role in the realm of matrix reordering.

 A key step in the RCM algorithm is to convert the matrix into an undirected graph and construct level structure in the undirected graph. In this step, building a level structure different nodes as a starting node will give results with very different structures. This difference will directly affect the quality of the results of the RCM algorithm.Therefore, the RCM algorithm typically requires a heuristic algorithm to determine an appropriate starting point in order to construct the level structure. This issue is commonly referred to as the starting node problem of the RCM algorithm.Several scholars have proposed different algorithms to solve this problem. \cite{arany1983another}\cite{reid1999ordering}\cite{kaveh1986ordering}\cite{kaveh1991connectivity}\cite{pachl1984finding}\cite{smyth1985}\cite{gonzaga2018evaluation} Among them, Kaveh's algorithms \cite{kaveh1986ordering}\cite{kaveh1991connectivity}\cite{kaveh2000ordering} and Pachl's \cite{pachl1984finding} algorithm are the more classical examples.
 
Among the algorithms the George-Liu (GL) algorithm currently enjoys widespread popularity.Its implementation has been incorporated into several prominent programming libraries, including the Boost library for C++, \cite{boost2017} the NetworkX library for Python, \cite{networkx_website} and MATLAB's \texttt{symcm} function. \cite{mathworks2018} In their work, Gonzaga de Oliveira and de Abreu \cite{gonzaga2018evaluation} compared the GL algorithm with several algorithms mentioned earlier. The results indicated that the GL algorithm is the best in terms of balancing both time efficiency and the quality of the results. Its versatility across matrices of various sizes and configurations contributes to achieving optimal reordering outcomes, making the combination of the GL algorithm with the RCM algorithm a prevalent preprocessing approach in matrix analysis.

In our paper, we innovatively propose addressing the starting node problem of the RCM algorithm by simultaneously considering the eccentricity and width of the nodes. Based on this idea, we introduce the Bi-Criteria pseudo-peripheral Node Finder (BNF) algorithm. By integrating this algorithm with the RCM algorithm, we develop the RCM++ algorithm. A series of experimental results strongly indicate that RCM++ achieves higher quality results in nearly the same amount of time compared to the current mainstream GL+RCM approach.

\section{Contribution}

The contributions of this paper can be summarized in the following three aspects:

\textbullet Provided an open-source, categorized dataset for the RCM starting node problem.

\textbullet Designed the BNF algorithm and combined it with RCM to develop the RCM++ algorithm.

\textbullet Designed and conducted experiments demonstrating that RCM++ is a superior algorithm compared to the RCM implementations in major software packages.
\section{Preliminary}

This section introduces the fundamental symbols and definitions used throughout the paper. Section 3.1 covers essential mathematical symbols and definitions, while Section 3.2 explains specific terms used in this paper.

\subsection{Basic}

\textbf{Bandwidth :} The bandwidth of a matrix \( A \) is defined as the maximum distance from the main diagonal to the farthest non-zero element in the matrix. Formally, the bandwidth \( b \) of a matrix \( A \) of size \( n \times n \) is given by:

\[
b = \max_{i,j \mid A_{ij} \neq 0} |i - j|
\]

where \( A_{ij} \) denotes the element in the \( i \)-th row and \( j \)-th column of the matrix \( A \).

\noindent \textbf{Profile:} The profile of a matrix \( A \) is defined as the sum of the distances of the non-zero elements from the main diagonal. Formally, for a matrix \( A \) of size \( n \times n \), the profile \( P(A) \) is given by:

\[
P(A) = \sum_{i=1}^{n} (i - \min\{ j \mid A_{ij} \neq 0 \})
\]

where \( A_{ij} \) denotes the element in the \( i \)-th row and \( j \)-th column of the matrix \( A \). The profile measures the sparsity structure of the matrix by accounting for the positions of its non-zero elements relative to the main diagonal.

\noindent \textbf{Degree of a node: } The degree of a node \( v \) in a graph \( G = (V, E) \) is defined as the number of edges incident to \( v \). Formally, the degree of a node \( v \) is denoted as \(\text{deg}(v)\). This metric indicates how many direct connections a node has to other nodes in the graph.

\noindent \textbf{Eccentricity of a node: } The eccentricity of a node \( v \) in a graph \( G = (V, E) \) is defined as the greatest distance from \( v \) to any other node in the graph. Formally, the eccentricity \( \epsilon(v) \) of a node \( v \) is given by:

\[
\epsilon(v) = \max_{u \in V} d(v, u)
\]

where \( d(v, u) \) represents the shortest path distance between nodes \( v \) and \( u \). The eccentricity of a node can also be referred to as its depth, denoted as \(\text{depth}(v)\).

\noindent \textbf{Level Structure:} Given a graph \( G = (V, E) \) and a starting vertex \( v \in V \), the level structure \( L(v) \) based on Breadth-First Search (BFS) is defined as follows:

\begin{itemize}
    \item \( L_0(v) = \{ v \} \)
    \item For \( i \geq 0 \), \( L_{i+1}(v) = \{ u \in V \mid \exists w \in L_i(v) \text{ and } (w, u) \in E \text{ and } u \notin \bigcup_{j=0}^{i} L_j(v) \} \)
\end{itemize}

Here, \( L_i(v) \) represents the set of vertices at distance \( i \) from the starting vertex \( v \).

\noindent \textbf{Width of a node:} The width of a node \( r \) in a graph \( G \) is defined as the maximum width among all levels in its rooted level structure \( L(r) \). Formally, the width \( w(r) \) is given by:

\[
w(r) = \max_{1 \leq i \leq h} \left\{|l_i(r)|\right\}
\]

where \( |l_i(r)| \) represents the number of nodes in the \( i \)-th level of \( L(r) \), and \( h \) is the total number of levels in \( L(r) \).

\subsection{Terminology Explanation}

\noindent \textbf{Node quality:} The quality of a node refers to the quality of the reordered matrix obtained after applying the Reverse Cuthill-McKee (RCM) algorithm starting from this node. Considering that the goal of the RCM algorithm is to minimize the bandwidth and profile of a sparse matrix as much as possible, it is generally believed that the smaller the bandwidth and profile of the reordered matrix, the higher the quality of the result. Therefore, the quality of a node is directly related to the bandwidth and profile of the reordered matrix.

\noindent \textbf{GL\_P:} refers to an array used to store all the nodes traversed by the GL algorithm. The GL algorithm is a heuristic algorithm that, based on certain rules, starts from a random node in an undirected graph, traverses a subset of nodes, and returns the last node as the result. We will explain the GL algorithm in detail later.

\noindent \textbf{RCM++: }In this paper, RCM++ refers to the combination of BNF and RCM for matrix preprocessing. Other combinations will be indicated by their abbreviations, such as GL\_RCM.

\noindent \textbf{Relative difference: }The relative difference refers to the proportional difference between two data points. The relative difference between A and B can be expressed as:

\begin{equation}
\text{Relative Difference} = \frac{A - B}{A}
\end{equation}

This metric can be used to measure the extent of increase or decrease of B relative to A.
\section{Related work}

\noindent \textbf{GL algorithm: }The GL algorithm is currently the most widely used and most effective algorithm among all algorithms addressing the starting node problem in RCM. The goal of the BNF algorithm proposed in this paper is to find an algorithm that has a similar runtime to the GL algorithm but produces nodes of higher quality.
Its design is based on the following two observations:

\begin{observation}
In the starting node problem in RCM , it is generally accepted that nodes with greater eccentricity are of higher quality. Specifically, for two nodes \( v \) and \( u \), if \( \text{eccentricity}(v) >\text{eccentricity}(u) \), then \( v \) is considered to be a better candidate.
\end{observation}

\begin{observation}
In the starting node problem in RCM, it is commonly assumed that the smaller the degree of the node, the higher its quality. That is, for two nodes \( v \) and \( u \), if \( \deg(v) < \deg(u) \), then \( v \) is regarded as a better candidate.
\end{observation}

Listing 1 specifically illustrates the process of the GL algorithm. Essentially, the GL algorithm attempts to find the node in the graph that has the greatest eccentricity and the smallest degree. It begins by selecting a random node and performing a breadth-first search to compute levels. Iteratively, it selects the node with the minimum degree from the deepest level, updates the breadth-first search levels, and repeats until no node at a deeper level is found. The algorithm ultimately returns  an appropriate starting node for the RCM algorithm.

\begin{lstlisting}[caption={George-Liu algorithm},label=alg:GeorgeLiu,captionpos=t,float,abovecaptionskip=-\medskipamount,mathescape=true]
Input: graph $G=(V, E)$
Output: a node $v \in V$

$v \gets$ RandomNode($V$)
$L(v) \gets$ BreadthFirstSearch($v$)
%BreadthFirstSearch($v$) means Construct L(v) starting from v.

while True do
    $u \gets$ MinimumDegreeNode($L_{\epsilon(v)}(v)$)
    $L(u) \gets$ BreadthFirstSearch($u$) % Count the eccentricity of $u$
    if $ \epsilon(u) > \epsilon(v)$ then
        $v \gets u$
        $L(v) \gets L(u)$
    else
        break 
    endif
endwhile

return $u$
\end{lstlisting}

\noindent \textbf{Minimum degree algorithm : }
The MIND (minimum degree algorithm) is also one of the most widely used algorithms in the RCM starting node problem. The algorithm is quite simple: it iterates through all the nodes in the undirected graph converted from the matrix and returns the one with the smallest degree as the result.Thanks to its simple design, it has high implementability. As a result, this method is widely used in various software packages.

The issue with this method is its high randomness, as there may be multiple nodes in the graph with the smallest degree simultaneously. This problem becomes more pronounced as the matrix size increases. Generally, the quality of the nodes obtained using this method is inferior to that of the GL algorithm.

\section{BNF algorithm}

The BNF algorithm, short for Bi-Criteria Node Finder algorithm, is proposed in this paper, as a modification of the GL algorithm. The core idea is based on Observation 3, introducing width as an additional evaluation metric into the GL algorithm. This allows the new algorithm to consider both the width and eccentricity of nodes while searching for suitable starting nodes.

\begin{observation}
In the starting node problem in RCM , it is generally accepted that nodes with smaller width are of higher quality. Specifically, for two nodes \( v \) and \( u \), if \( \text{width}(v) >\text{width}(u) \), then \( u \) is considered to be a better candidate.
\end{observation}

Listing 2 details the process of the BNF algorithm. Specifically, the BNF algorithm essentially performs the GL algorithm while recording the width of all nodes traversed by the GL algorithm. Finally, the result is determined based on the width of the nodes, rather than directly outputting the last node as the result like the GL algorithm. Although this idea may appear straightforward, it is actually the result of extensive experimentation with various potential methods to incorporate width into the GL algorithm. The process involved considerable effort and was not a trivial task.

This method has two significant theoretical advantages:
\begin{enumerate}
    \item The runtime of BNF is almost identical to that of the GL algorithm because the width of a node can be calculated using the same level structure as used for calculating the eccentricity of the same node.
    \item It can achieve higher quality results than the GL algorithm because, in many cases, the result produced by the GL algorithm is not the highest quality node among all the nodes it traverses.
\end{enumerate}

\begin{figure*}[!htb]
  \centering
  \begin{subfigure}[t]{0.3\linewidth}
    \centering
    \includegraphics[width=\linewidth]{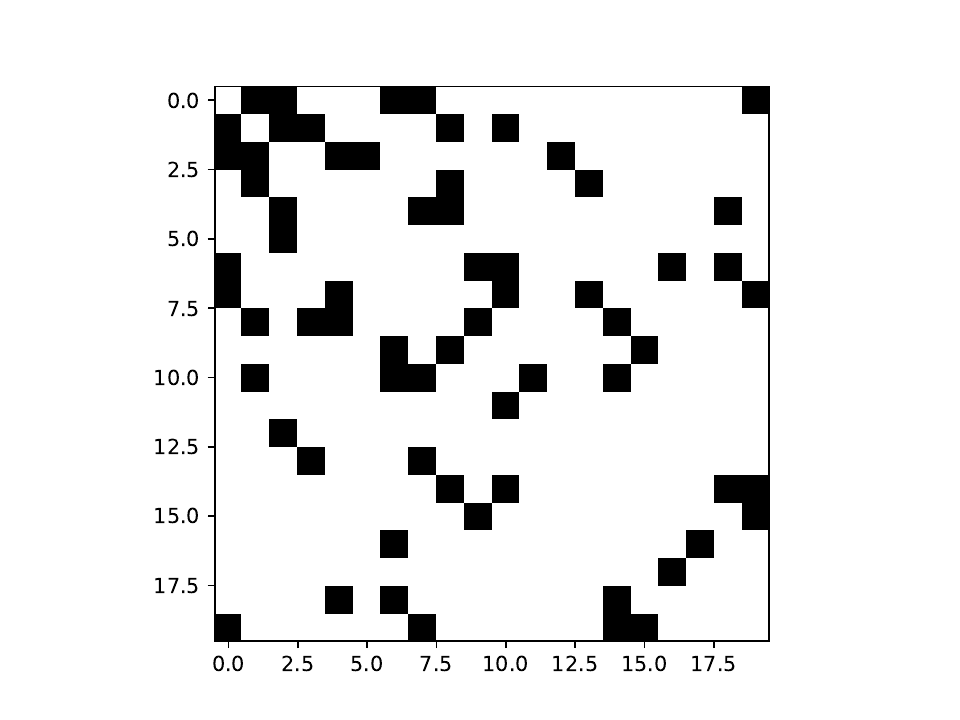} 
    \caption{Origin matrix.}
    \label{fig:origin_matrix}
  \end{subfigure}\hfill
  \begin{subfigure}[t]{0.3\linewidth}
    \centering
    \includegraphics[width=\linewidth]{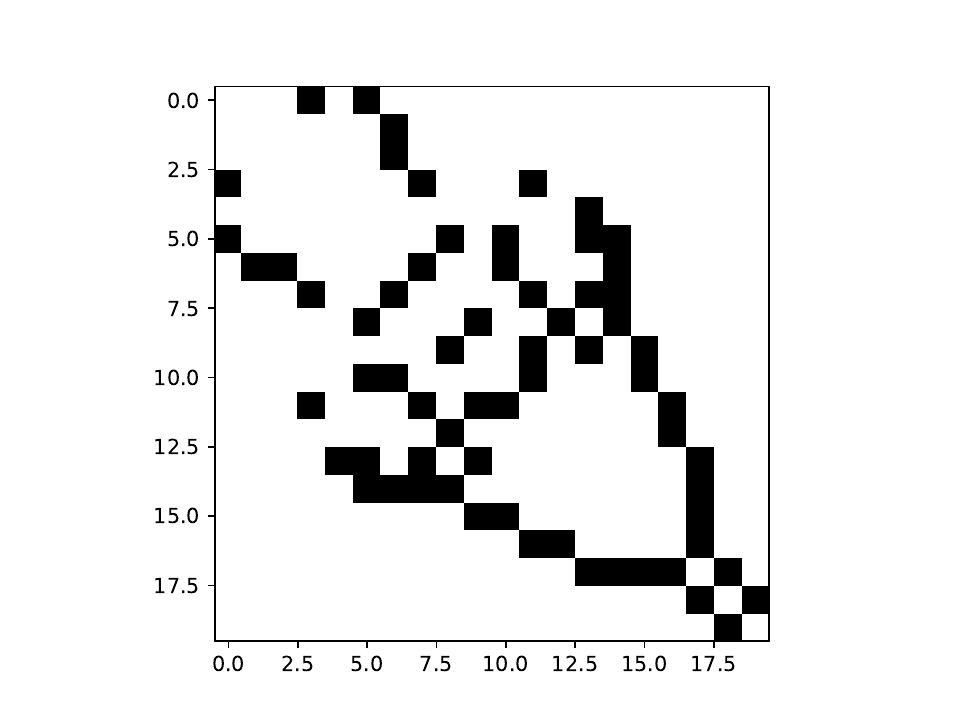}
    \caption{Matrix by RCM. bandwidth=9, profile=94.}
    \label{fig:matrix_by_rcm}
  \end{subfigure}\hfill
  \begin{subfigure}[t]{0.3\linewidth}
    \centering
    \includegraphics[width=\linewidth]{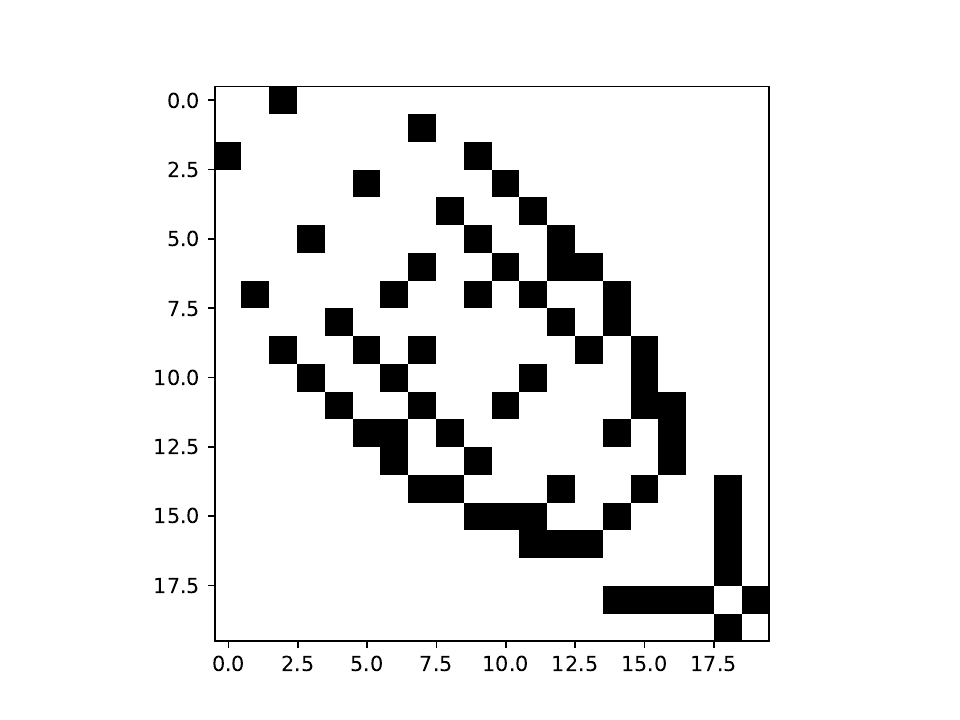}
    \caption{Matrix by RCM++. bandwidth=7, profile=88.}
    \label{fig:matrix_by_rcmpp}
  \end{subfigure}

  \begin{subfigure}[t]{0.3\linewidth}
    \centering
    \includegraphics[width=\linewidth]{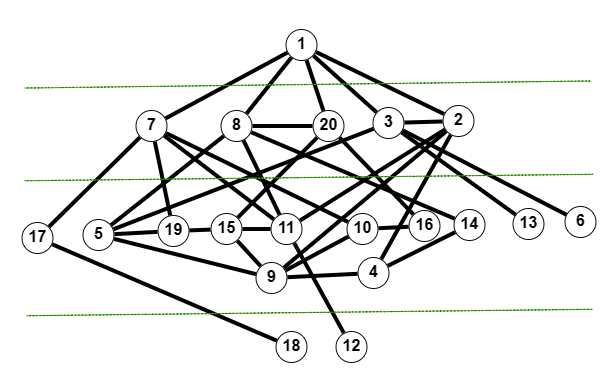}
    \caption{Level structure starts at 1.}
    \label{fig:level_structure_1}
  \end{subfigure}\hfill
  \begin{subfigure}[t]{0.3\linewidth}
    \centering
    \includegraphics[width=\linewidth]{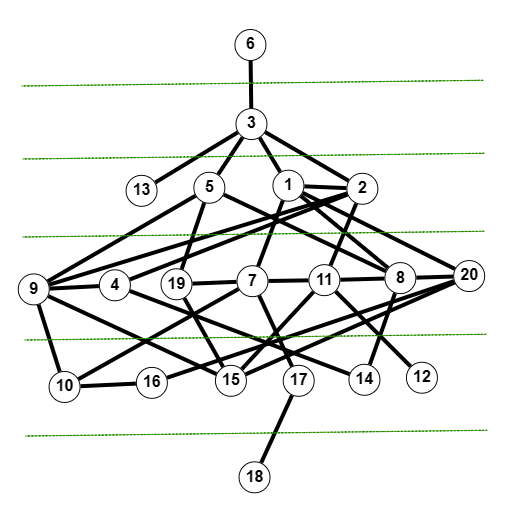}
    \caption{Level structure starts at 6. The green lines help distinguish levels.}
    \label{fig:level_structure_6}
  \end{subfigure}\hfill
  \begin{subfigure}[t]{0.3\linewidth}
    \centering
    \includegraphics[width=\linewidth]{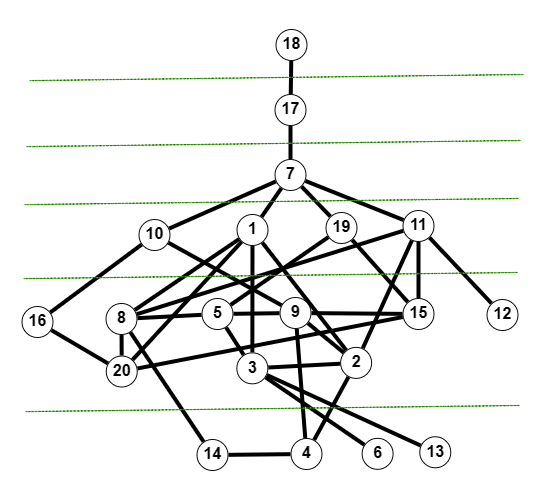}
    \caption{Level structure starts at 18.}
    \label{fig:level_structure_18}
  \end{subfigure}

  \caption{A small matrix is used as an example to illustrate the superiority of the RCM++ algorithm over the GL\_RCM algorithm. The green lines distinguish levels.Different algorithms were chosen for different results, GL chose node 18 while BNF chose node 6.}
  \label{fig:matrix_comparison}
\end{figure*}

We provide an example to illustrate why BNF can identify higher quality nodes compared to the GL algorithm. Suppose I want to apply both the GL algorithm and the BNF algorithm to the sparse matrix in Figure~\ref{fig:matrix_comparison}(a), starting from node 1. Under this assumption, both the GL algorithm and the BNF algorithm will traverse the nodes 1, 12, 6, and 18 in sequence.That means \[
BNF\_P = GL\_P = [1, 12, 6, 18]
\]

Figure~\ref{fig:matrix_comparison}(e) and Figure~\ref{fig:matrix_comparison}(f) respectively show the level structures rooted at node 6 and node 18. As can be seen from the figures, the eccentricitys of these two nodes are equal.That is:
\[
eccentricity(GL\_P[4]) = eccentricity(GL\_P[3]))=5
\]
Based on the rules of the GL algorithm, since the final two nodes have the same eccentricity, the GL algorithm will stop and return the last traversed node, which is node 18, as the final result. In contrast, the BNF algorithm takes an additional step by sequentially recording the widths of the traversed nodes and identifying the one with the smallest width. This enables the BNF algorithm to notice that:

\[
\text{width}(6) = \min_{v \in \text{GL\_P}} \text{width}(v),
\]

Therefore, BNF outputs node 6 as the final result.Figure~\ref{fig:matrix_comparison}(b) shows the result of the RCM algorithm starting from node 18, while Figure~\ref{fig:matrix_comparison}(c) shows the result of the RCM algorithm starting from node 6. It can be seen that the reordered matrix obtained by RCM with the BNF algorithm has a smaller bandwidth and profile. This indicates that the node selected by BNF is of higher quality than the one selected by GL.

The reason the BNF algorithm can improve the GL algorithm in this example lies in a potential drawback of the GL algorithm. When the GL algorithm stops, the last two nodes it traverses always have the same eccentricity. Since the GL algorithm relies solely on eccentricity for its decisions, it cannot differentiate between nodes with the same eccentricity. Therefore, introducing width as an additional criterion is necessary and can enhance the performance of the GL algorithm in many cases.

\begin{lstlisting}[caption={Bi-Criteria Node Finder},label=alg:BiCriteria,captionpos=t,float,abovecaptionskip=-\medskipamount,mathescape=true]
Input: graph $G=(V, E)$
Output: a node $v \in V$

$r \gets$ Randomnode($V$)
recordWidth $\gets \infty$
recordedNode $\gets 0$
$L(r) \gets$ BreadthFirstSearch($r$)
% UpdateWidth is a function that will count the width and 
% eccentricity of input level structure and update it
UpdateWidth($L(r)$, recordWidth, recordedNode) 

while True do
    $x \gets$ MinimumDegreenode($L_{\epsilon(v)}(r)$)
    $L(x) \gets$ BreadthFirstSearch($x$)
    UpdateWidth($L(x)$, recordWidth, recordedNode) 
    if $\epsilon(x) > \epsilon(r)$ then
        $r \gets x$
        $L(r) \gets L(x)$
    else
        break
    endif
endwhile

return recordedNode
\end{lstlisting}

\section{Experiment}

In this experimental section, we compare the proposed RCM++ algorithm with other algorithms regarding speed, result quality, and practicality. Section 6.1 covers the preliminary preparations, including comparison objects, hardware, and software environments. Section 6.2 introduces the dataset used and explains the experimental design. Section 6.3 presents a comparison of result quality between RCM++ and other algorithms. Section 6.4 compares the execution time of BNF with other algorithms. Finally, Section 6.5 demonstrates the performance of the RCM++ algorithm in solving equations compared to other algorithms.

\subsection{Experimental preparation}
Our experimental hardware configuration is as follows: The processor is a 13th Gen Intel Core i7-13800H with a clock speed of 2.50 GHz. The system is equipped with 32 GB of RAM. The operating system is 64-bit, running on an x64-based processor. 

For the software environment, we used Python 3.11.5. The functions used for comparison in the experiments include the GL algorithm and the Minimum Degree algorithm from Python's NetworkX library version 3.1. The choice of these two algorithms for comparison is mainly due to their prevalence in major software packages and their high practicality. Other algorithms are not widely used, often due to issues with result quality or time consumption. Gonzaga de Oliveira and de Abreu \cite{gonzaga2018evaluation} discussed several algorithms for solving the RCM starting node problem and concluded that the GL algorithm is currently the most practically valuable.

\subsection{Dataset and Experimental Design}

The experimental analyses carried out in this study involved 1049 symmetric matrices from the SuiteSparse Matrix Collection \cite{SparseMatrixCollection}. These matrices vary significantly in scale, ranging from 82 to 17,467,046 nonzeros, providing a robust basis for evaluating. Furthermore, a subset of positive definite matrices was selected from this collection to facilitate the experiments focused on solving equations.

The dataset used in our experiments includes matrices of varying sizes. This significant variation makes absolute time and bandwidth reduction metrics less meaningful. For instance, reducing the computation time by one second for a \(1000 \times 1000\) matrix is vastly different from achieving the same reduction for a \(100000 \times 100000\) matrix.

A more reasonable approach is to present the relative differences between algorithms rather than their absolute differences. In the experimental section of this paper, to better demonstrate the differences between our algorithm and the popular GL algorithm, we used the widely applied MIND algorithm as a baseline and adopted the concept of relative difference. While this methodology is relatively uncommon in experimental studies, we believe it is a rational design choice given the unique characteristics of our dataset.

The large number and diverse structures of the matrices often introduce significant noise in the experimental results. Therefore, we frequently use exponential smoothing in the experimental section.

\begin{figure}[!htbp]
\centering
\includegraphics[width=12cm]{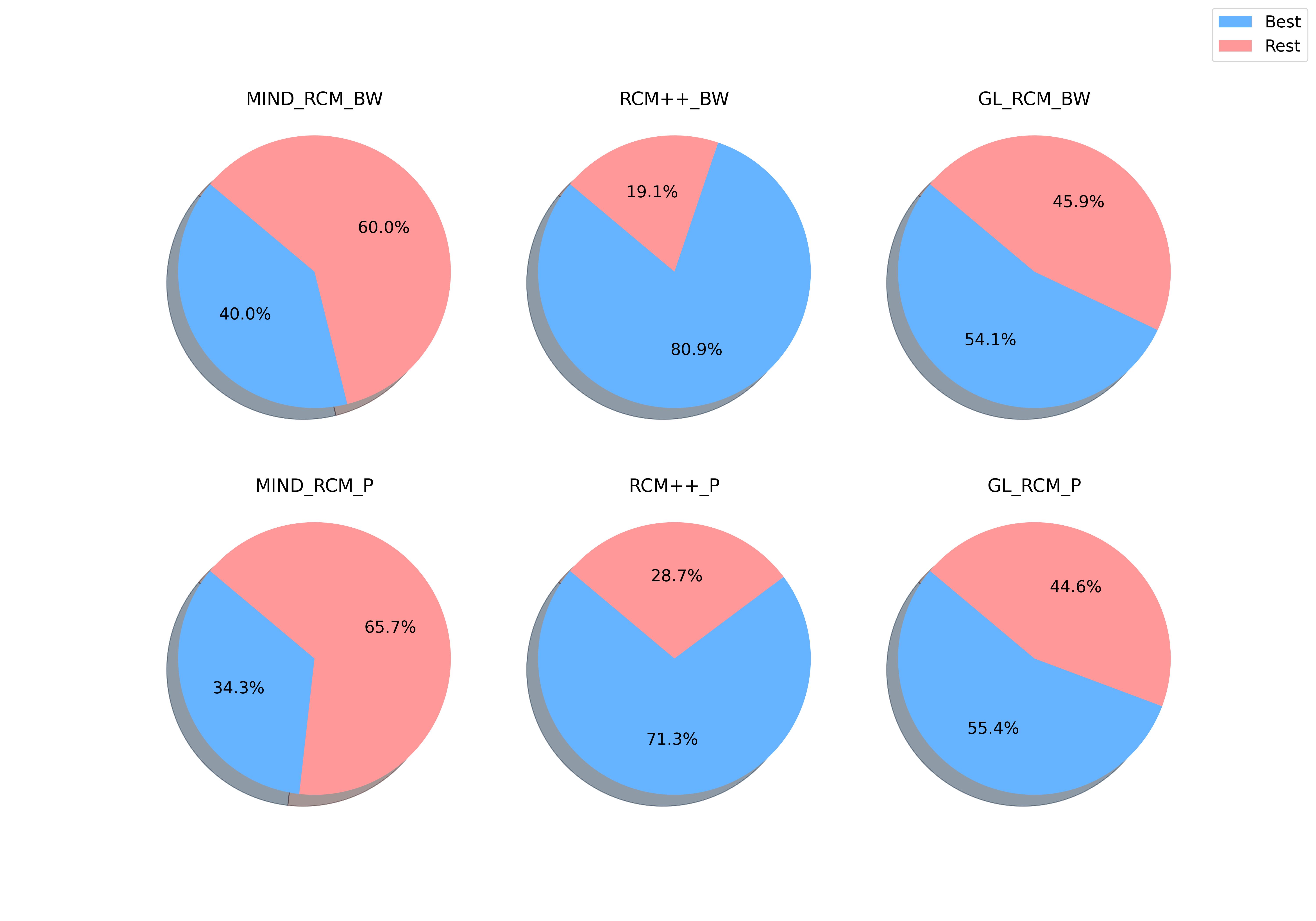}
\caption{This figure presents the proportion of optimal results selected by each method out of the total results.The suffix BW represents the use of bandwidth as the evaluation metric, while the suffix P indicates the use of profile as the evaluation metric. It can be observed that RCM++ consistently selects a higher number of optimal results in terms of both bandwidth and profile.}
\label{fig:speed}
\end{figure}

\subsection{Comparison of Algorithm Result Quality}

In this section, we use three methods—RCM++, GL\_RCM, and MIND\_RCM to reorder the matrices in the dataset. We then compare the reordered matrices based on their bandwidth and profile. These two metrics measure the maximum distance and the total distance of non-zero elements from the diagonal in a sparse matrix. For the algorithm results, smaller bandwidth and profile values indicate higher quality outcomes. To minimize random errors, both the GL algorithm and the BNF algorithm start from the same initial node.

Figure 3 shows the proportion of times each algorithm achieved the optimal result. From both bandwidth and profile perspectives, the BNF algorithm outperforms the others, achieving the optimal solution in most cases. In contrast, the MIND algorithm has the fewest instances of obtaining the optimal solution. The proportion of optimal results achieved by the BNF algorithm is significantly different from the other methods.

\begin{figure}[!htb]
\centering
\includegraphics[width=13cm]{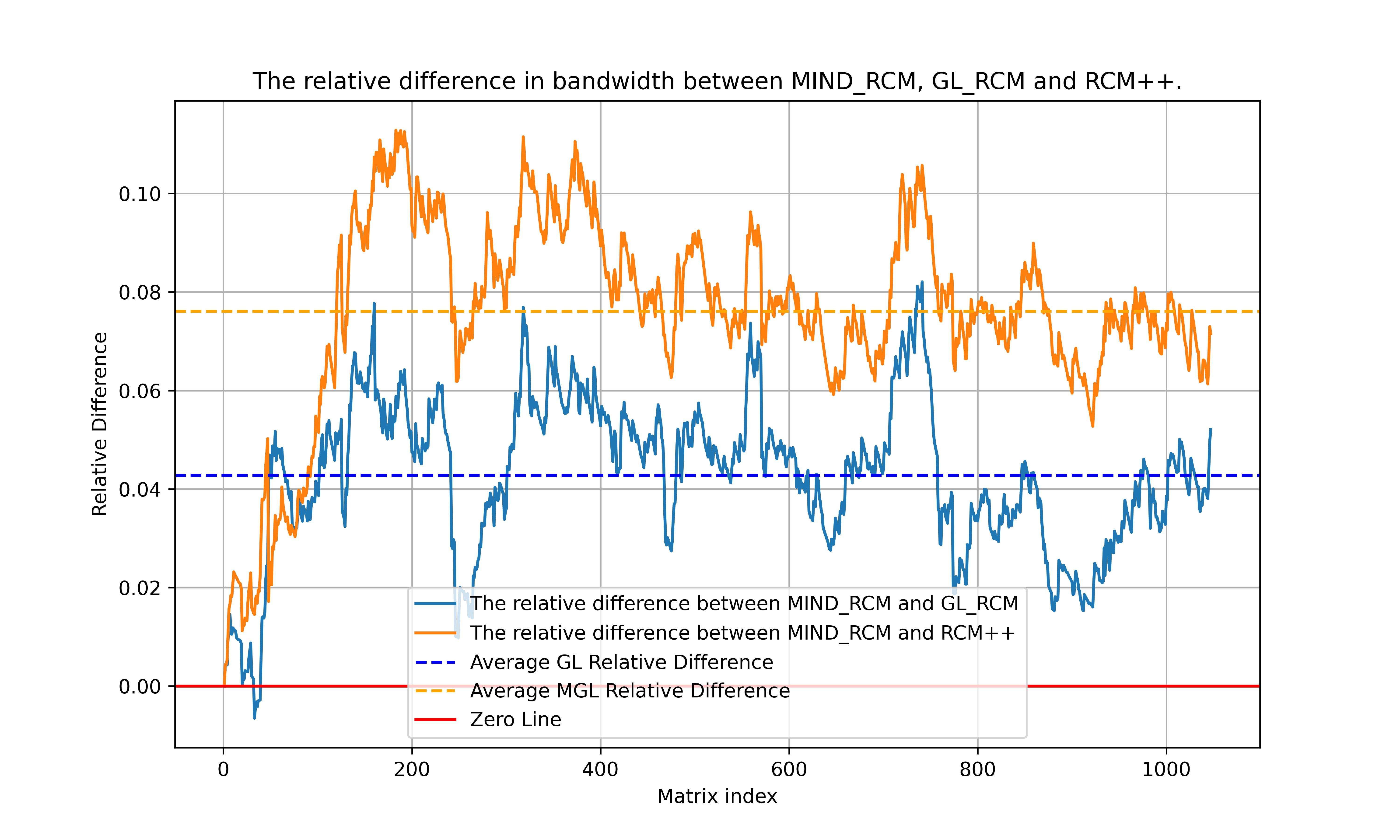}
\includegraphics[width=13cm]{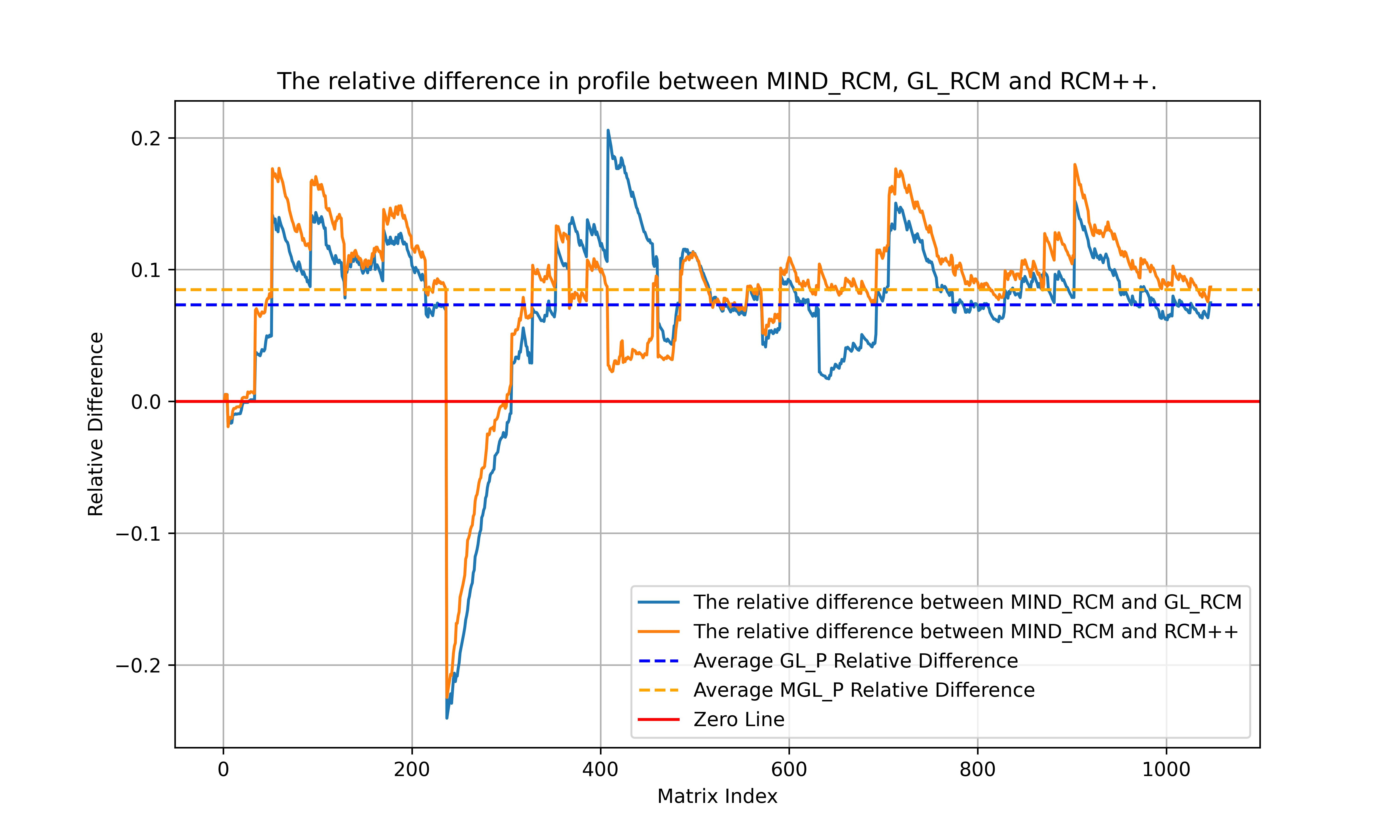}
\caption{The relative differences in bandwidth and profile between the result of MIND\_RCM algorithm compared to the result of GL\_RCM and RCM++ algorithms are shown. The X-axis represents the index of matrices sorted by size in ascending order. The data in the figure has been smoothed using exponential smoothing with a span of 100.A higher relative difference indicates a stronger ability to reduce bandwidth and profile. As shown, RCM++ almost always achieves better results.}
\label{fig:speed}
\end{figure}
\vspace{-5mm}
Figure 4 illustrates the relative differences between the MIND\_RCM algorithm compared to the GL\_RCM and RCM++ algorithms It is evident that both GL\_RCM and RCM++ generally outperform MIND\_RCM and exhibit similar trends. In most cases, RCM++ achieved better experimental results than both GL\_RCM and MIND\_RCM, with the relative differences remaining consistent regardless of matrix size. This indicates that RCM++ is suitable for matrices of varying sizes.

The instances where RCM++ underperforms compared to other algorithm will be discussed in detail in the subsequent error analysis section.

\subsection{Comparison of Running Times for Different Algorithms}

In this section, we ran the GL, BNF, and MIND algorithms 100 times each on all matrices in the dataset and measured their running times. Since the main difference between RCM++, GL\_RCM, and MIND\_RCM lies in the algorithm for finding the starting node, this experiment directly assesses the efficiency of RCM++. Considering these algorithms have very short running times and are minimally affected by matrix structure and size, we present the absolute running times for all three algorithms.

Figure 5 shows the running times of the three algorithms. As can be seen, the GL algorithm is the fastest, while the running times of MIND and BNF are similar.

\begin{figure}[!htb]
\centering
\includegraphics[width=0.8\linewidth]{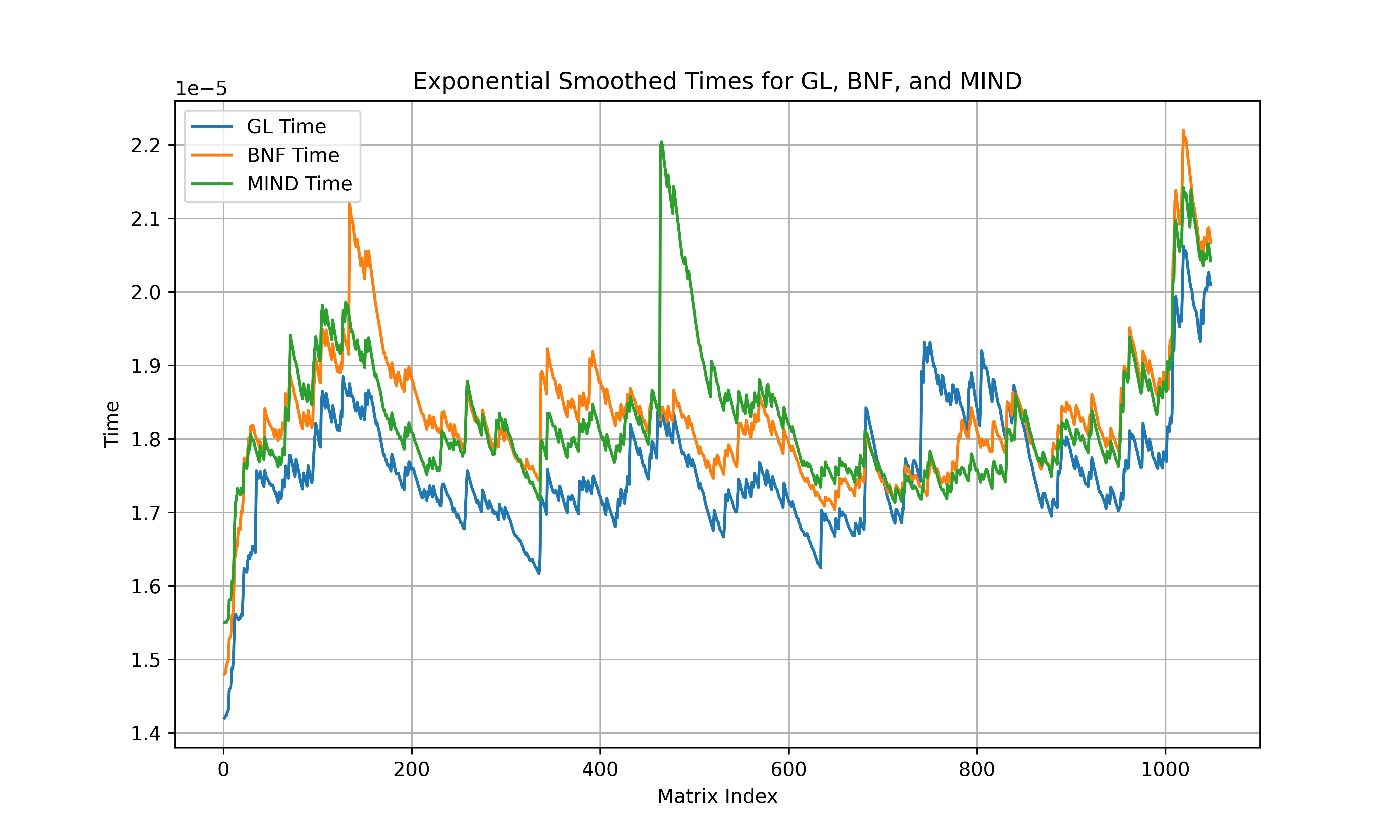}
\caption{The running times of GL, BNF, and MIND over 100 iterations are shown, with the X-axis representing the index of matrices sorted by size. The data has also been smoothed using exponential smoothing with a span of 100. It can be observed that the running time of BNF is similar to that of MIND, while GL has the fastest running time. The differences between them are minimal.}
\label{fig:time_line_graph}
\end{figure}

In terms of running time, BNF has a slight disadvantage. However, given its extremely short running time, this difference is almost negligible for larger matrix sizes. For instance, with the hood matrix, a single run of the BNF algorithm takes approximately 1.6e-07 seconds, while solving the matrix equation takes around 0.76 seconds.

\subsection{Application to solving matrix equations}

The RCM algorithm finds its applications in areas such as graph theory, parallel computing, and accelerating sparse matrix-vector multiplication. However, its simplest and most common use is in solving systems of equations.

Chan and George \cite{chan1980linear} have underscored the significance of the RCM algorithm, especially in its application to matrix equations solved via the Cholesky decomposition\cite{higham2009cholesky}. The RCM algorithm is notably effective in minimizing the envelope of the permutation matrix associated with the Cholesky decomposition, thereby enhancing the efficiency of solving matrix equations.

We extracted non-singular matrices from the original dataset and reordered them using RCM++, GL\_RCM, and MIND\_RCM algorithms. Finally, we calculated and compared the speeds of solving the reordered matrices using Cholesky decomposition for each algorithm.

\begin{figure}[!htb]
\centering
\includegraphics[width=0.8\linewidth]{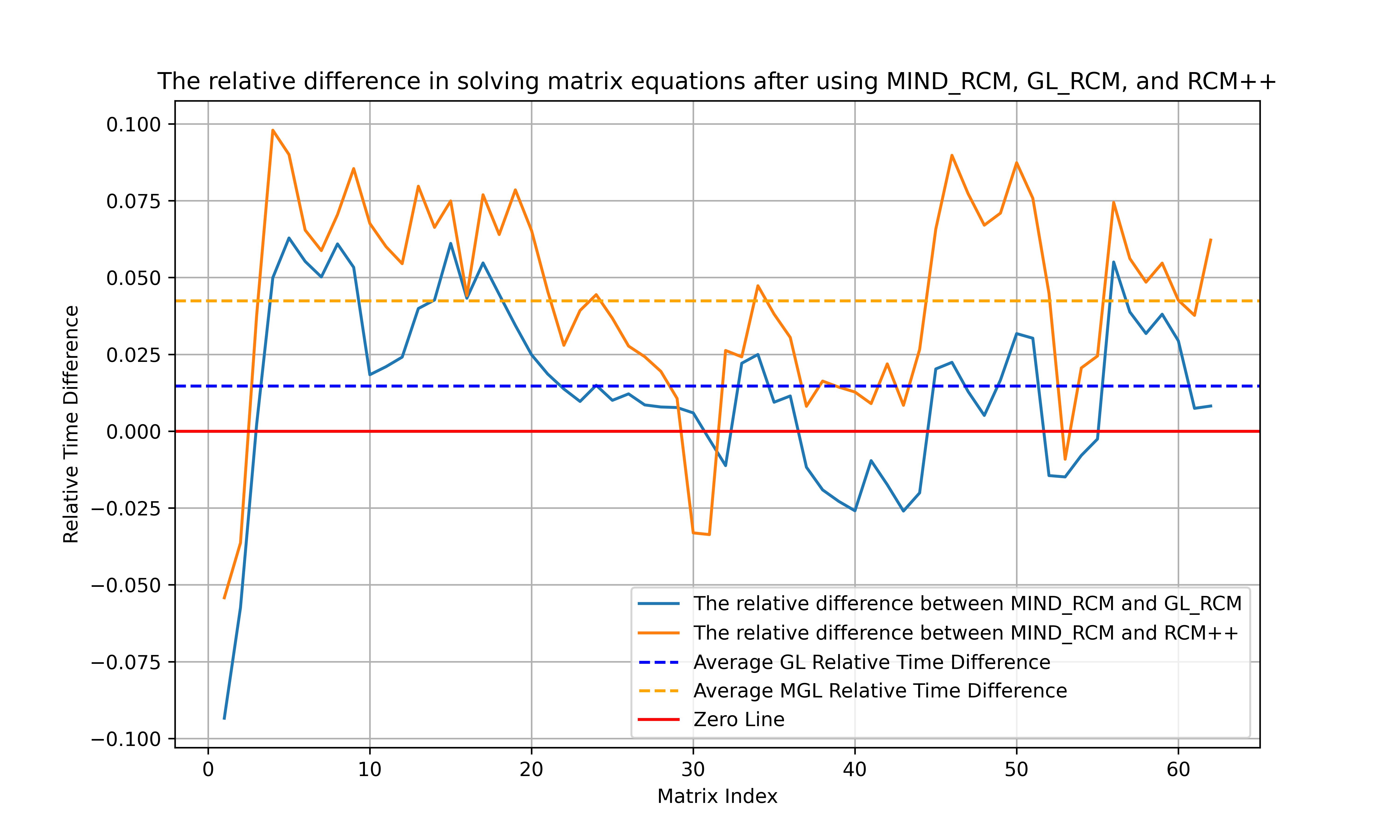}
\caption{This figure shows the relative time differences in solving matrix equations using Cholesky decomposition after reordering the matrices with GL\_RCM, MIND\_RCM, and RCM++. A higher relative difference indicates a more significant acceleration effect for solving matrix equations compared to MIND\_RCM. RCM++ clearly demonstrates the best performance, and its effectiveness is less influenced by matrix size. The data in this figure has been smoothed using exponential smoothing with a span of 10.}
\label{fig:time_line_graph}
\end{figure}

Figure 6 shows the time taken to solve equations after reordering with RCM++ compared to the other algorithms. It is evident that RCM++ consistently outperforms GL and MIND, regardless of matrix size.

\subsection{Error Analysis}
In the relative difference curve of the Profile shown in Figure 4, RCM++ performs worse than GL\_RCM for matrices indexed from 400 to 500. However, in the bandwidth curve, BNF achieves better results for the same subset of matrices.

Upon studying this subset, we found they have relatively random structures. BNF selects nodes with the same eccentricity as GL but with smaller width. According to Observation 3, these nodes are generally considered higher quality. However, for these matrices, using such nodes as starting points for RCM results in smaller bandwidth but larger profile. These discrepancies result from the randomness in matrix structure.

\section{Conclusion and Future work}

In our research, we proposed the BNF algorithm to address the starting node problem of RCM. This algorithm considers both width and eccentricity attributes during the search for suitable nodes. We used publicly available matrix databases to establish a test dataset and designed experiments based on three criteria: speed, quality, and practicality. Through these experiments, we conducted a detailed comparison of RCM++ with GL\_RCM and MIND\_RCM from the Python NetworkX library. The results indicate that RCM++ effectively improves result quality with minimal additional time cost and significantly enhances solving matrix equations in practical scenarios.

Furthermore, BNF holds potential for application in other matrix reordering algorithms, such as the Solan \cite{https://doi.org/10.1002/nme.1620281111} and GPS \cite{gibbs1976} algorithms. In future work, we will study the application of the BNF algorithm in other matrix reordering algorithms and integrate the results to develop a comprehensive software package.

\nocite{cuthill1969reducing}
\nocite{george1981solution}
\nocite{kaveh1986ordering}
\nocite{gonzaga2018evaluation}
\nocite{boost2017}
\nocite{mathworks2018}
\nocite{networkx_website}
\nocite{kaveh2000ordering}
\nocite{kaveh1991connectivity}
\nocite{pachl1984finding}
\nocite{reid1999ordering}
\nocite{smyth1985}
\nocite{gibbs1976}
\nocite{pyutils2023}
\nocite{arany1983another}
\nocite{davis2011university}
\nocite{liu1976comparative}
\nocite{chinn1982bandwidth}
\nocite{chan1980linear}
\nocite{higham2009cholesky}
\nocite{10.1145/3581784.3607046}
\nocite{alappat2020recursive}
\nocite{SparseMatrixCollection}
\bibliography{RCM++}

\begin{appendices}

\section{Supplementary experimental data charts}
\begin{table*}[!htb]
\centering
\caption{Shows specific experimental data on 15 matrices that have been used in the course of research by other scholars \cite{gonzaga2018evaluation}.On experiments with small samples, RCM++ performs better than large samples. This reflects the potential of RCM++ for practical applications.}
\begin{tabular}{|l|c|c|c|c|c|c|}
\hline
\textbf{Matrix Name} & \multicolumn{3}{c|}{\textbf{\emph{bandwidth}}} & \multicolumn{3}{c|}{\textbf{\emph{profile}}} \\
\hline
& \textbf{GL\_RCM} & \textbf{RCM++} & \textbf{Diff \%} & \textbf{GL\_RCM} & \textbf{RCM++} & \textbf{Diff\%} \\
\hline

1138 bus & 132 & 111 & 15.91\% & 44912 & 41190 & 8.29\%  \\
494 bus & 63 & 63 & 0.00\% & 10661 & 10661 & 0.00\%  \\
662 bus & 118 & 88 & 25.4\% & 28722 & 20223 & 29.5\% \\
685 bus & 102 & 66 & 35.2\% & 25834 & 18879 & 26.9\%  \\
ash85 & 13 & 10 & 23.08\% & 619 & 589 & 4.85\%  \\
bcspwr01 & 8 & 6 & 25.00\% & 123 & 108 & 12.20\%  \\
bcspwr06 & 121 & 100 & 17.36\% & 62956 & 54382 & 13.62\%  \\
bcspwr09 & 133 & 137 & -3.01\% & 79277 & 82385 & -3.92\%  \\
bcsstk04 & 54 & 54 & 0.00\% & 3965 & 3965 & 0.00\%  \\
bcsstk05 & 26 & 25 & 3.85\% & 2314 & 2280 & 1.47\%  \\
dwt 503 & 58 & 65 & -12.07\% & 15530 & 14822 & 4.56\%  \\
gearbox & 6339 & 4239 & 33.13\% & 327006240 & 278201886 & 14.92\%  \\
msdoor & 8414 & 6558 & 22.06\% & 1393116060 & 1243344049 & 10.75\%  \\
delaunay n21 & 11057 & 9283 & 16.04\% & 593691094 & 372553004 & 37.25\%  \\
wave & 8405 & 7395 & 12.02\% & 727397699 & 648899237 & 10.79\%  \\

\hline
\end{tabular}
\end{table*}

\end{appendices}

\end{document}